# Nonreciprocal singularities dominated by the dissipative photon-magnon coupling in non-Hermitian systems

Yongzhang Shi[1], Chi Zhang[1], Zhenhui Hao[1], Changjun Jiang[1], C. K. Ong[1,2], Ke Xia[3], Guozhi Chai[1*]

[1]*Key Laboratory for Magnetism and Magnetic Materials of the Ministry of Education, School of Physical Science and Technology, Lanzhou University, Lanzhou, 730000, People's Republic of China,*

[2]*Department of Physics, Xiamen University Malaysia, Jalan Sunsuria, Bandar Sunsuria, 43900, Sepang, Selangor, Malaysia.*

[3]*Department of Physics, Southeast University, Jiangsu 211189, China*

**Abstract**

We investigated the magnon-photon coupling in an open cavity magnonic system, which leads to two different nonreciprocal singularities dominated by the dissipative coupling. One type of singularity is the exceptional point, which is just on the exceptional surface in parameter space. The other type of singularity is the bound state in the continuum discovered in the level-attraction-like coupling, which is above the exceptional surface. In experiment, we realized the two different singularities with nonreciprocity and selectivity in an open cavity magnonic system with suitable dissipation rating. Our results can be understood well with the pseudo-Hermitian theory of magnon-polariton system.

---

* Correspondence email address: chaigzh@lzu.edu.cn

*Introduction*.—Photon-magnon coupling is a subject of cavity quantum electrodynamics won great attention in the last decade since the theoretical work of Soykal and Flatté in 2010 [1]. After that, the photon–magnon coupling using the bulk cavity has been realized experimentally [2-4] and also realized in some other systems [5-23]. In recent years, level attraction in cavity magnonics due to the new type of dissipative coupling has been discovered [24]. Then, many studies about dissipative coupling have been achieved in vary systems [25-33]. Among which, a controllable transition between level attraction and level repulsion has been realized by using a designed planar microstrip-cross-junction cavity [27]. In the dissipative coupling systems, two types of singularities in cavity magnonics, exceptional points (EPs) and bound states in the continuum (BICs), have attracted much attention in photon-magnon coupling system.

EPs are square-root singularities appearing in non-Hermitian systems. The surface consisting of EPs in parameter space is called the exceptional surface (ES). When the system reaches the EP condition, both the eigenstates and eigenvalues of the system coalesce. In a two-state system, this manifests as a degeneration of the two branches of dispersion. EPs have been studied in a wide range of systems, such as optics [34-37], microwave [38-41] and also in magnonics [30,42,43]. On the other hand, the BICs are different kind of singularities which were originally predicted by von Neumann and Wigner [44]. At BIC, a perfectly confined state is bound in the continuum, which can't radiate away, means zero linewidth in experiment. It has led to a wide range of applications in many systems including photonics [45-49], magnonics [26,28,30] and acoustics [50,51].

Beyond the broad interest to realize both singularities, nonreciprocal operations are important in controlling the signal transmission and detection in electromagnetic waves from microwave to optical frequency [52-56]. In previous works, the unconventional singularity has been observed under purely dissipative coupling in an anti-parity-time symmetric cavity magnonics system [30], and the mirror symmetric nonreciprocity has also been realized in an open cavity magnonic system [26]. However, to achieve the singularities with nonreciprocal behaviors is still challenging. In this letter, we design an open cavity magnonic system with suitable dissipation rating not only to achieve both types of singularities (EPs and BICs), but also to achieve the nonreciprocal and selectivable singularities. In this design, only one EP could be realized with microwave propagating in certain direction.

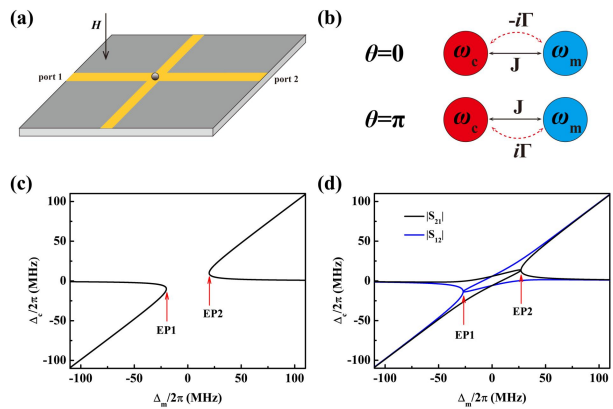

FIG. 1. (a) Sketch of the experimental setup. The planar cavity is placed in the *xy* plane, with an external magnetic field *H* applied along the *z* direction (perpendicular to the planar cavity). The spectroscopy is measured with a vector network analyzer through two ports. (b) Schematic diagram of the photon-magnon coupling. It shows the difference between the forward and backward transmissions. (c) and (d) Two cases when the EPs appear, for two EPs in (e) and one EP in (f).

*System and model.*—Based on the coupling theory [26], the photon mode (cavity mode) $\omega_c$ and the magnon mode $\omega_m$ should interact with each other through the coherent coupling ($J$) and the dissipative coupling ($\Gamma$) caused by radiation damping, as shown in Fig. 1(b). The non-Hermitian Hamiltonian of such a cavity magnonic system can be expressed as [26]:

$$\hat{H}/\hbar = (\omega_c - i\beta)\hat{a}^+\hat{a} + (\omega_m - i\alpha)\hat{b}^+\hat{b} + (J - i\Gamma e^{i\theta})(\hat{a}^+\hat{b} + \hat{b}^+\hat{a}) . \qquad (1)$$

Here, $\alpha$ and $\beta$ are the intrinsic damping rates of the magnon and cavity modes, respectively. $\hat{a}(\hat{a}^+)$ and $\hat{b}(\hat{b}^+)$ are the photon and magnon annihilation (creation) operators, respectively. $J$ is the coherent coupling strength, and $\Gamma$ is the dissipative coupling strength. $\theta=0$ and $\pi$ for $S_{21}$ and $S_{12}$, respectively. It can be proved that the Hamiltonian is η-pseudo-Hermitian and satisfies the symmetry *CPT* [43], where *C* mimics charge conjugation symmetry, *P*, and *T* are not the same as parity ($\mathcal{P}$) and time ($\mathcal{T}$) symmetries. The calculated transmission coefficient of the system using input-output theory can be represented as [26]:

$$S_{21(12)} = 1 + \frac{\kappa}{i(\omega-\omega_c)-(\kappa+\beta)+\frac{-[iJ+\Gamma e^{i\theta}]^2}{i(\omega-\omega_m)-(\alpha+\gamma)}} . \qquad (2)$$

Here, $\kappa$ and $\gamma$ are the external damping rates of the cavity and magnon modes, respectively. A term of $-2iJ\Gamma e^{i\theta}$ in Eq. (2) produces nonreciprocity on account of the interference of $J$ and $\Gamma$.

The eigenvalues of Eq.(1) can be calculated as:

$$\tilde{\omega}_{\pm} = \frac{1}{2}\left[\omega_c + \omega_m - i(\beta + \alpha) \pm \sqrt{[(\omega_m - i\alpha) - (\omega_c - i\beta)]^2 + 4(J - ie^{i\theta}\Gamma)^2}\right]$$

$$= \frac{1}{2}\left[\omega_c + \omega_m - i(\beta + \alpha) \pm \sqrt{\mathcal{G}}\right] \quad (3)$$

$$\text{Re}(\mathcal{G}) = (\omega_m - \omega_c)^2 - (\beta - \alpha)^2 + 4(J^2 - \Gamma^2)$$

$$\text{Im}(\mathcal{G}) = 2(\omega_m - \omega_c)(\beta - \alpha) \pm 8J\Gamma$$

corresponding to two hybridized modes. The EPs emerge at a threshold value when Re($\mathcal{G}$)= Im($\mathcal{G}$) = 0. At the setting of EPs, Re($\tilde{\omega}_{\pm}$)=($\omega_c$+$\omega_m$)/2, Im($\tilde{\omega}_{\pm}$)=-($\alpha$+$\beta$)/2, that is, both the resonance frequency and the intrinsic damping rate are the average values of the two modes (the photon mode and the magnon mode). The ES consisting of EPs is shown in Fig. 2. On the ES in Fig. 2(a), Re($\mathcal{G}$)=0, two hybridized modes intersect with each other, i.e., two branches of the dispersion degenerate. There are two cases with different number of EPs existed. One is like the purely dissipative coupling, in which, there are two EPs in the dispersion as shown in Fig. 1(d) as reported by many work [24,25,30]. While there is only one EP in one dispersion for $S_{21}$ or $S_{12}$ in another case, as shown in Fig. 1(e), which should be discussed in details. Above the ES, Re($\mathcal{G}$)>0, it corresponds to η-pseudo-Hermiticity (*CPT* symmetry) unbroken region. Two hybridized modes are separated with each other. There is no intersection in the two branches of the dispersion. When *J*>Γ, the coupling of level-repulsion-like dominated by coherent coupling existed which is similar with the results reported by our previous work [17]. As comparison, when *J*<Γ, the coupling of level-attraction-like dominated by dissipative coupling which will be discussed below. The BICs emerge when the imaginary parts of Eq. (3) (Im($\tilde{\omega}_{\pm}$)) equals zero. The theoretical intrinsic damping rates Im($\tilde{\omega}_{\pm}$) go to zero. Extreme narrow peaks will appear under the setting of BICs in the experimental results, indicating that the damping is extreme low.

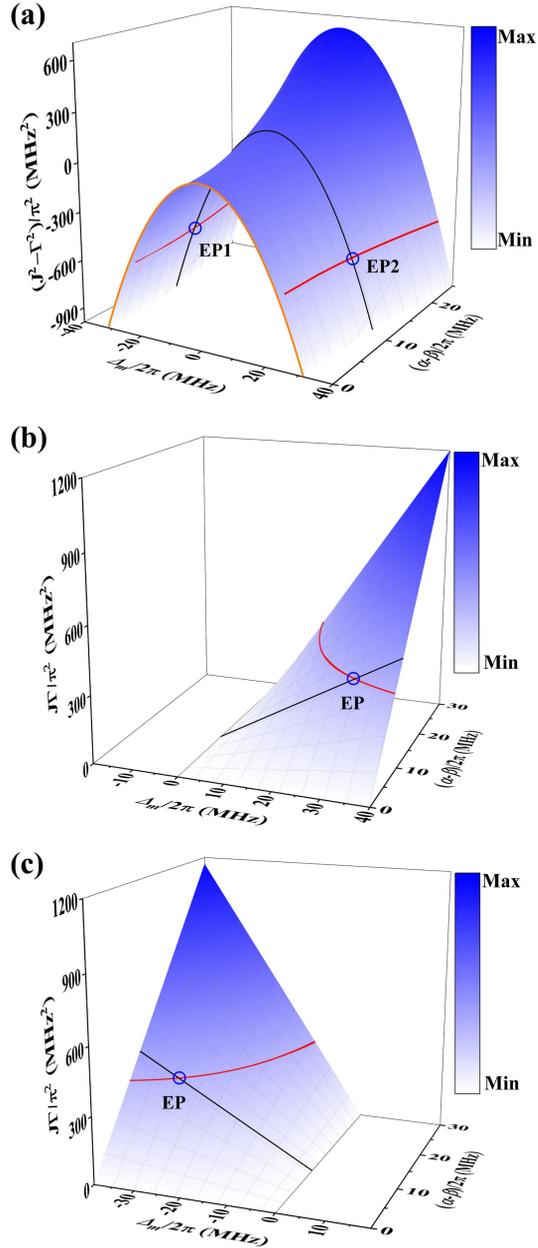

FIG. 2. The ES in the parameter space as a function of $\Delta_m$ and $(\alpha-\beta)$. (a)The ES when $\mathrm{Re}(\mathcal{G}) = 0$. (b)The ES when $\mathrm{Im}(\mathcal{G}) = 0$ for $S_{21}$. (c)The ES when $\mathrm{Im}(\mathcal{G}) = 0$ for $S_{12}$.

The experimental setup is schematically shown in Fig. 1(a). The microstrip system forming an cross-line open cavity is placed in the $xy$ plane providing the cavity photon mode. The side length of the rectangle plate is 40 mm. The device consists of a cross-line microwave circuit with a dimension as $1.15 \times 40$ mm$^2$. A YIG sphere with 0.45 mm diameter is placed just in the middle of the microwave circuit. The microwave signals and the transmission spectrum are measured with a vector network analyzer (VNA). An external static magnetic field is applied perpendicular to the cavity plane to control the frequency of magnon mode $f_m$, which can be calculated by the Kittel equation ($f_m = \omega_m/2\pi = \gamma_e H$, $\gamma_e$ is the gyromagnetic ratio).

*Experimental results*.—Figures 3(a) and 3(b) show, respectively, the density mapping images of the transmission coefficient $S_{21}$ and $S_{12}$ as a function of the frequency detuning ($\Delta_c = \omega - \omega_c$) and field detuning ($\Delta_m = \omega_m - \omega_c$). The coupling strength can be obtained by fitting the experimental results with the eigenvalues of Eq. (1) as the following formula:

$$\Delta\widetilde{\omega}_\pm = \frac{1}{2}\left[\Delta\omega_m - i(\beta+\alpha) \pm \sqrt{[\Delta\omega_m + i(\beta-\alpha)]^2 + 4(J - ie^{i\theta}\Gamma)^2}\right], \quad (4)$$

where the real parts and the imaginary parts of $\Delta\widetilde{\omega}_\pm$ (Re($\Delta\widetilde{\omega}_\pm$) and Im($\Delta\widetilde{\omega}_\pm$)) are the frequency detuning and the intrinsic damping rates of the coupled resonances respectively. Re($\Delta\widetilde{\omega}_\pm$) give the dispersions of two hybridized modes. The resonant frequency and damping rates of the cavity mode are $\omega_c/2\pi = 2.52$ GHz, $\kappa/2\pi = 875.4$ MHz and $\beta/2\pi = 6.4$ MHz are determined by fitting the the cavity transmission spectrum. The black dashed lines are the fitting dispersion curves in Fig. 3, and figures 3(c) and 3(d) are calculation results for Figs. 3(a) and 3(b) respectively. The calculation results agrees very well with the experimental data. Not only the dispersions but also the line width and the peak value are almost the same as the experimental data.

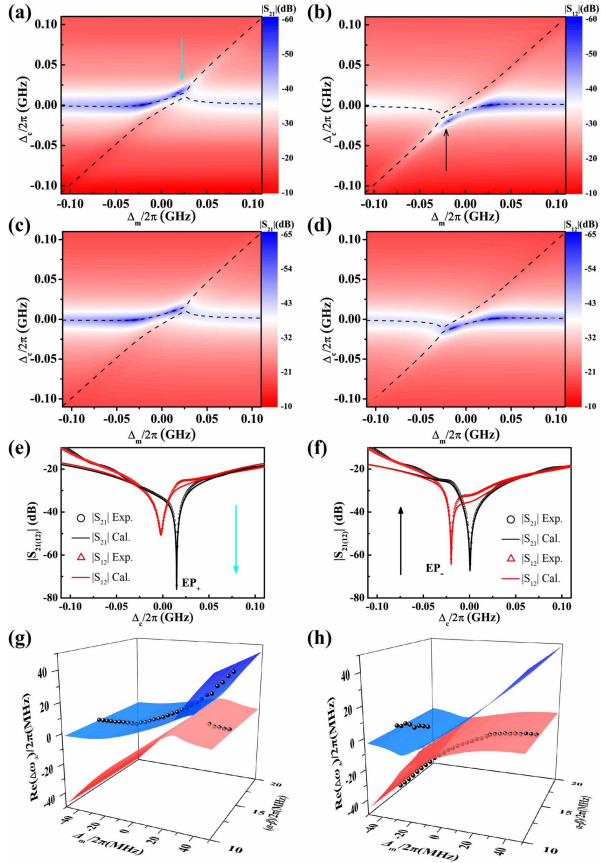

FIG. 3. Mapping image of the transmission coefficient $S_{21}$ [(a) and (c)] and $S_{12}$ [(b) and (d)] as a function of the frequency detuning ($\Delta_c$) and field detuning ($\Delta_m$). (a) and (b) are experimental data, (c) and (d) are calculation results. (e) and (f) $S_{21}$, $S_{12}$ spectra measured at the magnetic field marked by the blue arrow and black arrow, respectively. (g) and (h) The Riemann surfaces of $\text{Re}(\Delta\widetilde{\omega}_{\pm})$ in the parameter space as a function of $\Delta_m$ and ($\alpha$-$\beta$) for $S_{21}$ and $S_{12}$ respectively. The black points are experimental data.

The case of two EPs in the purely dissipative coupling have been reported by many work [24,25,30]. Our results show the case of only one EP in Fig. 3. In a certain S parameter mapping graph in Fig. 3(a) or 3(b), there is only one EP on the right or left from the center of the coupling as shown in blue arrow or black arrow, and the single EP is different for $S_{21}$ and $S_{12}$, which shows its mirror symmetric nonreciprocity. In the purely dissipative coupling, there are two EPs in the opposite frequency tuning from the center of the coupling as shown in Fig. 1(c). The coherent coupling strength $J = 0$ and it requires both $\text{Re}(\mathcal{G})=0$ and $\text{Im}(\mathcal{G})=0$ in Eq. (3) when EP appears. Im($\mathcal{G}$)=0, that is $\Delta_m = 0$ or $α-β = 0$. When $\Delta_m = 0$, Re($\mathcal{G}$)=0 has no solution; When $α-β = 0$, it correspond to the orange line in Fig. 2(a). There are always two points of intersection (EPs) regardless of the value of $\Gamma$.

If there is interaction between coherent coupling $J$ and dissipative coupling $\Gamma$ (neither $J$ nor $\Gamma$ equals 0), and $J$ is smaller than $\Gamma$, there are still two intersections as shown by the black and red lines in Fig. 2(a). It corresponds to the two EPs of the system. However, the ES calculated when the imaginary part Im($\mathcal{G}$) = 0 in Eq. (3) is shown in Figs. 2(b) and 2(c). The black line has only one intersection with the red line, corresponding to only one EP. When the EP appears, both Re($\mathcal{G}$)=0 and Im($\mathcal{G}$)=0 in Eq. (3) must be required, thus limiting the two EPs in Fig. 2(a) to only one EP. For the non-reciprocal $S_{21}$ and $S_{12}$ as shown in Figs. 2(b) and 2(c), the position of EP will be mirror symmetric. When a single EP appears in the system, the coordinate of EP in the parameter space can be calculated as $\Delta_m = 2\Gamma$ and $α-β = 2J$ corresponding to Fig. 2(b) or $\Delta_m = -2\Gamma$ and $α-β = 2J$ corresponding to Fig. 2(c). According to Eq. (4), the resonant frequency of EP is $\Delta_c = \Gamma$ corresponding to $S_{21}$ or $\Delta_c = -\Gamma$ corresponding to $S_{12}$ in Fig. 1(d). This is EP's property of mirror symmetric nonreciprocity as shown in Fig. 1(d).

It can be seen clearly from our resultts that there is only one EP in Fig. 3(a) marked by the blue arrow for $S_{21}$ or Fig. 3(b) marked by the black arrow for $S_{12}$. The EPs in $S_{21}$ and in $S_{12}$ are mirror symmetric. The fitting parameters in our results are $J/2\pi = 6.36$ MHz, $\Gamma/2\pi = 13.5$ MHz, $(α-β)/2\pi = 11.5$ MHz, which satisfying the coordinate of EP in parameter space mentioned above: $α-β = 2J$.

Figures 3(e) and 3(f) show the $S_{21}$ and $S_{21}$ spectra measured at the magnetic field marked by the blue arrow and black arrow in Figs. 3(a) and 3(b). The black peak for $S_{21}$ in Fig. 3(e) and the red peak for $S_{12}$ in Fig. 3(f) represent EPs, marked as $EP_+$ and $EP_-$ respectively. Two branches of the dispersion degenerate at the setting of EPs, corresponding to the only one narrow peak for $S_{21}$ or $S_{12}$ in Figs. 3(e) and 3(f). EP is selective to transmission coefficient according to the results. For $S_{21}$, EP only appears at $EP_+$, corresponding to $\Delta_m>0$, $\Delta_c>0$. For $S_{12}$, EP only appear at $EP_-$, corresponding to $\Delta_m<0$, $\Delta_c<0$. The Riemann surfaces of $Re(\Delta\tilde{\omega}_\pm)$ in the parameter space for $S_{21}$ and $S_{12}$ are plotted in Figs. 3(g) and 3(h). It can be clearly seen that both $Re(\Delta\tilde{\omega}_\pm)$ and $\Delta_m$ of the intersection between the upper and lower surfaces in Fig. 3(g) are larger than zero, which representing $EP_+$ for $S_{21}$ in Fig. 3(a) marked by the blue arrow. $Re(\Delta\tilde{\omega}_\pm)$ and $\Delta_m$ of the intersection in Fig. 3(h) are less than zero, which representing $EP_-$ for $S_{12}$ in Fig. 3(b) marked by the black arrow. It's highly consistent with the experimental results.

Our results show that singularities have flexible controllability. The appearance for EP can be controlled by the direction of transmission, the energy, the direction of the magnetic field. For example, EP appears at the position marked by the blue arrow in Fig. 3(a) for $S_{21}$. However, there is no EP in the same position in Fig. 3(b) for $S_{12}$. This means the appearance of EP can be controlled by the direction of transmission. There are two EPs in the lower left part (low energy state) and upper right part (high energy state) in Fig. 1(c). However there is only one EP in the lower left part or upper right part in Fig. 3(a) or 3(b), which means we can control the appearance of EP by adjusting the energy so that the system is in a high energy state or a low energy state. By changing the direction of the magnetic field, the strength of the upper branch and the down branch in $S_{21}$ and $S_{12}$ is interchanged, which means that the distribution of $S_{21}$ when the magnetic field is reversed is similar to that of Fig. 3(b) and $S_{12}$ is similar to that of Fig. 3(a). Therefore, we can also regulate EP by reversing the direction of the magnetic field.

The other type of singularity (BIC) is also discovered in our experimental results. Figures 4(a) and 4(b) show, respectively, the density mapping images of $S_{21}$ and $S_{12}$ as a function of $\Delta_c$ and $\Delta_m$. The fitting curves, plotted as the black dashed lines in Figs. 4(a) and 4(b), agree well with the measured dispersions. Figures 4(c) and 4(d) are calculation results for Figs. 4(a) and 4(b) respectively. The calculation results agrees very well with the experimental data. The obtained coupling strengths are $J/2\pi$ = 5.6 MHz and $\Gamma/2\pi$ = 16 MHz.

It can be seen from Figs. 4(a) and 4(b) that the dispersion is similar to that of the purely dissipative coupling [24-27]. But it is found that it is not true after fitting and calculation. And we find that two hybridized modes are separated with each other. There is no intersection in the two branches of the dispersion, that is there is no EP here. It is the coupling of level-attraction-like dominated by dissipative coupling when $J<\Gamma$. It is above the ES when Re($\mathcal{G}$)>0, which representing the system stays in the η-pseudo-Hermiticity (CPT symmetry) unbroken region. The Riemann surfaces of Re($\Delta\tilde{\omega}_\pm$) in the parameter space are plotted in Fig. 5(a). The upper and lower surfaces represent Re($\Delta\tilde{\omega}_+$) and Re($\Delta\tilde{\omega}_-$) respectively. EPs should be the intersection of the upper and lower surfaces. It can be clearly seen that there is no intersection in Fig. 5(a). It can also prove there is no EP in Figs. 4(a) and 4(b).

However, we can see the dark blue color in the very small region in Figs 4(a) and

4(b), which represents an extreme large peak value and an extreme small peak width.

It's similar to the property of BIC we mentioned above. Therefore, the S-parameter spectra at the magnetic field marked by the arrows in Figs. 4(a) and 4(b) are provided for detailed analysis.

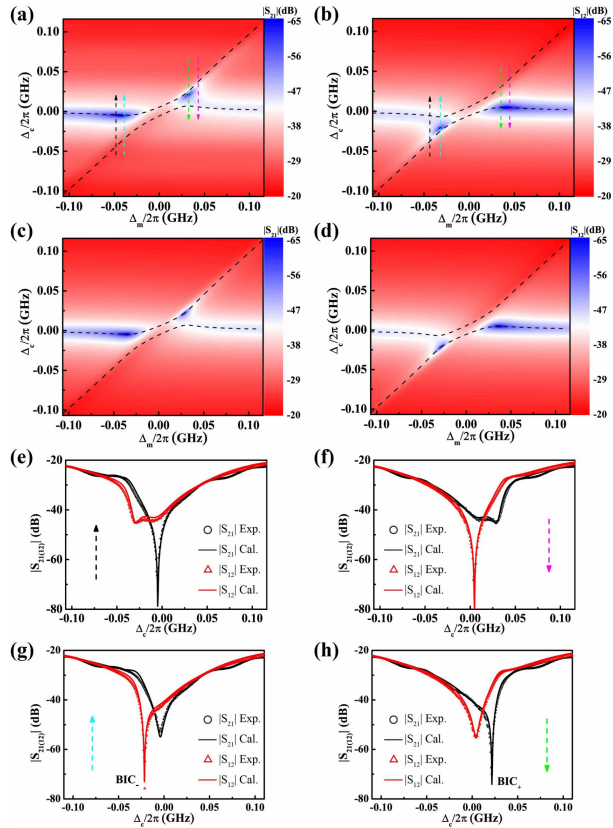

FIG. 4. Mapping image of the transmission coefficient $S_{21}$ [(a) and (c)] and $S_{12}$ [(b) and (d)] as a function of the frequency detuning ($\Delta_c$) and field detuning ($\Delta_m$). (a) and (b) are experimental data, (c) and (d) are calculation results. The magnetic field is applied in z direction. The black dash lines are the fitting curves according to Eq. (3). Black, magenta, blue and green arrows mark the bias fields at which the spectra plotted in (a)-(b) are measured. [(e)-(h)] $S_{21}$, $S_{12}$ spectra measured at the magnetic field marked by the black, magenta, blue and green arrows, respectively. The black and red solid lines are the fitting curves according to Eq. (2).

Figures 4(e)-4(h) show the $S_{21}$ and $S_{21}$ spectra measured at the magnetic field

marked by the black, magenta, blue and green arrow in Figs. 4(a) and 4(b), respectively. Large nonreciprocity can be found near the coupling area in Figs. 4(e)-4(h). Two peaks with large peak width can be seen in $S_{21}$ or $S_{12}$ in Figs. 4(e) and 4(f). However there is only one extreme narrow peak marked as BIC- in Fig. 4(g) and BIC+ in Fig. 4(h) respectively. The peak value can reach about -75 db, representing the complete block of the microwave transmission. The system reaches the condition of BICs at the extreme narrow peaks which imply the damping rates of the system go to zero.

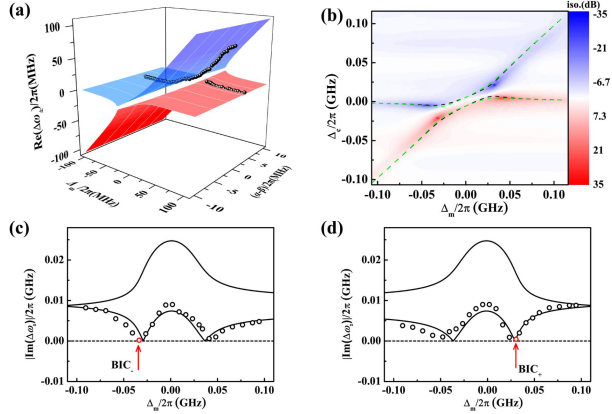

FIG. 5. (a) The isolation ratio as a function of $\Delta_c$ and $\Delta_m$. (b) The Riemann surfaces of Re($\Delta \tilde{\omega}_\pm$) in the parameter space as a function of $\Delta_m$ and ($\alpha$-$\beta$) for $S_{21}$. The black points are experimental data. (c) and (d) The imaginary parts of the eigenvalues (Im($\Delta \tilde{\omega}_\pm$)) as a function of $\Delta_m$ for $S_{21}$ and $S_{12}$ respectively. The black and red circles are experimental data.

The imaginary parts of the eigenvalues (Im($\Delta \tilde{\omega}_\pm$)) in Eq. (4) for $S_{21}$ and $S_{12}$ are plotted in Figs. 5(c) and 5(d) respectively. BICs should emerge when Im($\Delta \tilde{\omega}_\pm$) equals zero. BIC+ and BIC- are represented by red circles in Figs. 5(c) and 5(d), which are highly consistent with the theory. BIC is selective to transmission coefficient according to the results. BIC only appears at BIC+ for $S_{21}$, corresponding to $\Delta_m>0$, $\Delta_c>0$. For $S_{12}$, BIC only appears at BIC-, corresponding to $\Delta_m<0$, $\Delta_c<0$. We realized the nonreciprocal BICs in the coupling of level-attraction-like dominated by dissipative coupling.

Nonreciprocal transmissions can be summarized by comparing the Figs. 4(a) with 4(b). The strength of the upper branch in $S_{21}$ is about the same as the down branch in $S_{12}$, meanwhile the down branch in $S_{21}$ is similar to the upper branch in $S_{12}$, which implies a mirror symmetric nonreciprocity. By changing the direction of the magnetic field, the strength of the upper branch and the down branch in $S_{21}$ and $S_{12}$ is interchanged, which means that the mirror symmetric nonreciprocity can be reversed by reversing the field direction.

Next, we introduce the isolation ratio (iso. = $20 \times \log_{10}|S_{21}/S_{12}|$) , which shows the difference between the forward ($S_{21}$) and backward ($S_{12}$) transmission amplitudes [26], in order to quantify the nonreciprocity. Figure 5(b) show the isolation ratio as a function of $\Delta_m$ and $\Delta_c$. The fitting curves plotted as the black and green dashed lines in Fig. 5(b) agree well with the measured dispersions of the iso., which verify our fitting above. The extremums of the isolation ratios in the coupling area are larger than 40 dB, which shows the large nonreciprocity.

*Conclusion.*—In summary, we observe two different nonreciprocal singularities dominated by the dissipative coupling. One type of singularity is the EP, which is on the ES. The case of only one EP is also realized. The other type of singularity is BIC discovered in the coupling of level-attraction-like, which is above the ES. The two different singularities have nonreciprocity and selectivity. Our results verify the theory of non-Hermitian magnon-polariton system and and provide valuable ideas and experience for the subsequent study of non-Hermitian systems.

This work was supported by the National Natural Science Foundation of China(NSFC) (Nos. 51871117 and 12174165) and Natural Science Foundation of GanSu Province for Distinguished Young Scholars (No. 20JR10RA649).

**Conflict of Interest**

The authors have no conflicts to disclose.

**Data availability**

The data that support the findings of this study are available from the corresponding author upon reasonable request.